\documentclass[letterpaper]{article} 
\usepackage{aaai25}  
\usepackage{times}  
\usepackage{helvet}  
\usepackage{courier}  
\usepackage[hyphens]{url}  
\usepackage{graphicx} 
\usepackage{natbib}  
\usepackage{caption} 
\usepackage{algorithm}
\usepackage{algorithmic}
\usepackage{bm}
\usepackage{graphicx}
\usepackage{float}
\usepackage{subfigure}
\usepackage{amssymb}
\usepackage{amsmath}
\usepackage{amsthm}
\usepackage{booktabs}
\usepackage[switch]{lineno}
\usepackage{xcolor}
\usepackage{multirow}
\usepackage{bm}
\usepackage{newfloat}
\usepackage{listings}
\DeclareCaptionStyle{ruled}{labelfont=normalfont,labelsep=colon,strut=off} 
\floatstyle{ruled}
\newfloat{listing}{tb}{lst}{}
\floatname{listing}{Listing}
\title{Semantic-guided Masked Mutual Learning for Multi-modal Brain Tumor Segmentation with Arbitrary Missing Modalities}

\author {
    Guoyan Liang\textsuperscript{\rm 1}\equalcontrib,
    Qin Zhou\textsuperscript{\rm 2}\equalcontrib, Jingyuan Chen\textsuperscript{\rm 1}, \\
    Bingcang Huang\textsuperscript{\rm 3}, Kai Chen\textsuperscript{\rm 3}, Lin Gu\textsuperscript{\rm 4}, Zhe Wang\textsuperscript{\rm 2}\thanks{Corresponding Authors.}, Sai Wu\textsuperscript{\rm 1}\footnotemark[2], Chang Yao\textsuperscript{\rm 1}\footnotemark[2] \\
}

\affiliations {
    \textsuperscript{\rm 1}Zhejiang University, Hangzhou, China\\
    \textsuperscript{\rm 2}Department of Computer Science and Engineering, ECUST, China \\
    \textsuperscript{\rm 3}Gongli Hospital of Shanghai Pudong New Area   \\
    \textsuperscript{\rm 4}RIKEN AIP, The University of Tokyo  \\
    \{guoyanl, jingyuanchen, wusai, changy\}@zju.edu.cn, \{sunniezq, wangzhe\}@ecust.edu.cn
}

\usepackage{bibentry}
\usepackage{dsfont}

\begin{document}

\maketitle

\begin{abstract}
Malignant brain tumors have become an aggressive and dangerous disease that leads to death worldwide.
Multi-modal MRI data is crucial for accurate brain tumor segmentation, but missing modalities common in clinical practice can severely degrade the segmentation performance. While incomplete multi-modal learning methods attempt to address this, learning robust and discriminative features from arbitrary missing modalities remains challenging. To address this challenge, we propose a novel Semantic-guided Masked Mutual Learning (SMML) approach to distill robust and discriminative knowledge across diverse missing modality scenarios.
Specifically, we propose a novel dual-branch masked mutual learning scheme guided by Hierarchical Consistency Constraints (HCC) to ensure multi-level consistency, thereby enhancing mutual learning in incomplete multi-modal scenarios. The HCC framework comprises a pixel-level constraint that selects and exchanges reliable knowledge to guide the mutual learning process. Additionally, it includes a feature-level constraint that uncovers robust inter-sample and inter-class relational knowledge within the latent feature space. To further enhance multi-modal learning from missing modality data, we integrate a refinement network into each student branch. This network leverages semantic priors from the Segment Anything Model (SAM) to provide supplementary information, effectively complementing the masked mutual learning strategy in capturing auxiliary discriminative knowledge. Extensive experiments on three challenging brain tumor segmentation datasets demonstrate that our method significantly improves performance over state-of-the-art methods in diverse missing modality settings.
\end{abstract}

%

\section{Introduction}

The accurate segmentation of brain tumors is crucial for clinical assessment and surgical planning \cite{bakas2017advancing,yan2020neural}. In clinical practice, MRI scans across multiple imaging modalities, including T1-weighted (T1), contrast-enhanced T1-weighted (T1ce), T2-weighted (T2), and Fluid Attenuated Inversion Recovery (Flair) images, are combined to enable accurate brain tumor segmentation. However, it is not uncommon for one or more modalities to be absent due to image corruption, artifacts, acquisition protocols, allergies to contrast agents, or cost constraints \cite{qiu2023scratch}. Recently, incomplete multi-modal brain tumor segmentation has been frequently studied to deal with various missing modality scenarios \cite{ding2021rfnet,zhang2022mmformer,liu2023m3ae}. Despite significant advancements, it remains challenging to extract robust and discriminative multi-modal features when dealing with arbitrary missing modalities, which impacts segmentation performance. 

Based on the assumption that  brain tumor segmentation should yield consistent outcomes with the ground truth, irrespective of the modality combinations (e.g., T1$+$T2 or \text{Flair}$+$\text{T2}$+$\text{T1ce}) used for patient examination, we propose an innovative Semantic-guided Masked Mutual Learning (SMML) framework to integrate hierarchical consistency constraints along with auxiliary semantic priors from SAM, enhancing the robustness and discriminative ability of incomplete multi-modal learning. Our framework employs a masked dual-branch architecture to emulate a patient undergoing various examination scenarios with different missing modalities. Initially, full modality inputs are fed into both branches. We then randomly mask one or more modalities in each branch. The training process for each branch is supervised by the proposed hierarchical consistency constraints and the semantic prior derived from the SAM method.

To enhance the robustness of our dual-branch learning scheme with diverse missing modality inputs, we introduce hierarchical consistency constraints. These constraints operate on two levels: pixel-level and feature-level. The pixel-level bidirectional constraint enforces detailed consistency, ensuring that only reliable knowledge is exchanged during mutual learning. Simultaneously, to incorporate relational context into the distillation process, we leverage a feature-level constraint that explores the consistency of inter-sample and inter-class relationships between the two branches within the latent feature space.

To further improve the discriminative ability of each student branch, we introduce a semantic-guided refinement network. This network integrates auxiliary semantic information from SAM with initial predictions to generate refined segmentation outputs. Notably, the refinement network is employed solely during training, imposing no additional computational burden during inference. 

The overall contributions of our proposed method can be summarized as follows:
\begin{itemize}
    \item We propose a novel SMML framework to learn robust and discriminative knowledge across various missing modality setttings.
    \item We propose the hierarchical consistency constraints to enhance the robustness of the proposed SMML framework, where the pixel-level consistency selects and exchanges reliable knowledge across the dual branches, while the feature-level distillation guarantees the inter-sample and inter-class relational consistency. 
    \item We introduce a refinement network that incorporates auxiliary semantic knowledge from the well-established SAM, thereby enhancing the discriminative ability of each branch without incurring any computational burden during inference.
\end{itemize}

\section{Related Work}
\subsection{Multi-modal Brain Tumor Segmentation}
Complete multi-modal learning has markedly improved the accuracy of brain tumor segmentation from MRI scans \cite{wang2022uncertainty,10183842nnformer,10526382unetr}.
\cite{han2020trusted} design a trusted multi-view classifier to model the multi-modal uncertainties and fuse features via Dempster’s rule.
\cite{wang2022uncertainty} introduce an uncertainty-aware multi-modal learning model
through cross-modal random network prediction.

However, practical scenarios often involve missing modalities due to image corruption, allergies to contrast agents, or cost constraints \cite{liu2023m3ae}.
Therefore, many efforts have been made to accommodate the practical scenarios of missing modalities. While some methods address diverse missing modality settings by training modality-specific models \cite{liu2021face,karimijafarbigloo2023mmcformer}, these approaches overlook the valuable complementary information present across modalities. Recent efforts have focused on effectively learning representative multi-modal features that can adapt flexibly to diverse missing modality scenarios. 
\cite{ding2021rfnet} introduce a region-aware fusion module (RFM) to enhance feature representation by considering the diverse sensitivities of different modalities to tumor regions.
\cite{zhang2022mmformer} propose a transformer-based method that combines transformers and CNNs to learn modality-invariant representations.


\subsection{Mutual Learning}
Mutual Learning (ML) is a machine learning paradigm that involves the collaboration of multiple models to learn from each other.
This collaborative learning process allows the models to improve their performance by sharing knowledge. 
Mutual learning approaches, grounded in knowledge distillation, can be broadly categorized into response-based, feature-based, and relation-based methods. 
Response-based distillation \cite{hinton2015distilling} involves matching the soft labels produced by multiple models.
Feature-based distillation \cite{zagoruyko2016paying,huang2017like} focuses on aligning the pixel-wise or statistical feature distributions across different models, while relation-based distillation \cite{tung2019similarity,wei2023mmanet} matches the sample relations.

Mutual learning has been increasingly applied in scenarios involving incomplete multi-modal data.
\cite{li2021dynamic} propose to transfer the full-modal information from the teacher to guide the missing modality learning of the student branch. However, this approach necessitates a substantial volume of labeled full-modal data, which can be impractical to obtain in real-world applications.
Besides, dense pixel-level or feature-level distillation may fail unexpectedly if one or more crucial modalities are absent during incomplete multi-modal learning. 
To address this, some methods \cite{yang2022cross,wu2023extracting} focus on distill the structured knowledge through the relation-based or graph-based methods.
Despite notable advancements, there is a scarcity of mutual learning frameworks specifically designed to handle scenarios with arbitrary missing modality inputs. In this paper, we propose a novel semantic-guided masked mutual learning framework to enhance both the robustness and discriminative ability of mutual learning, thereby improving the performance of incomplete multi-modal learning with arbitrary missing modalities.

\section{Method}
\begin{figure*}[t]
\centering
\includegraphics[width=0.82\textwidth]{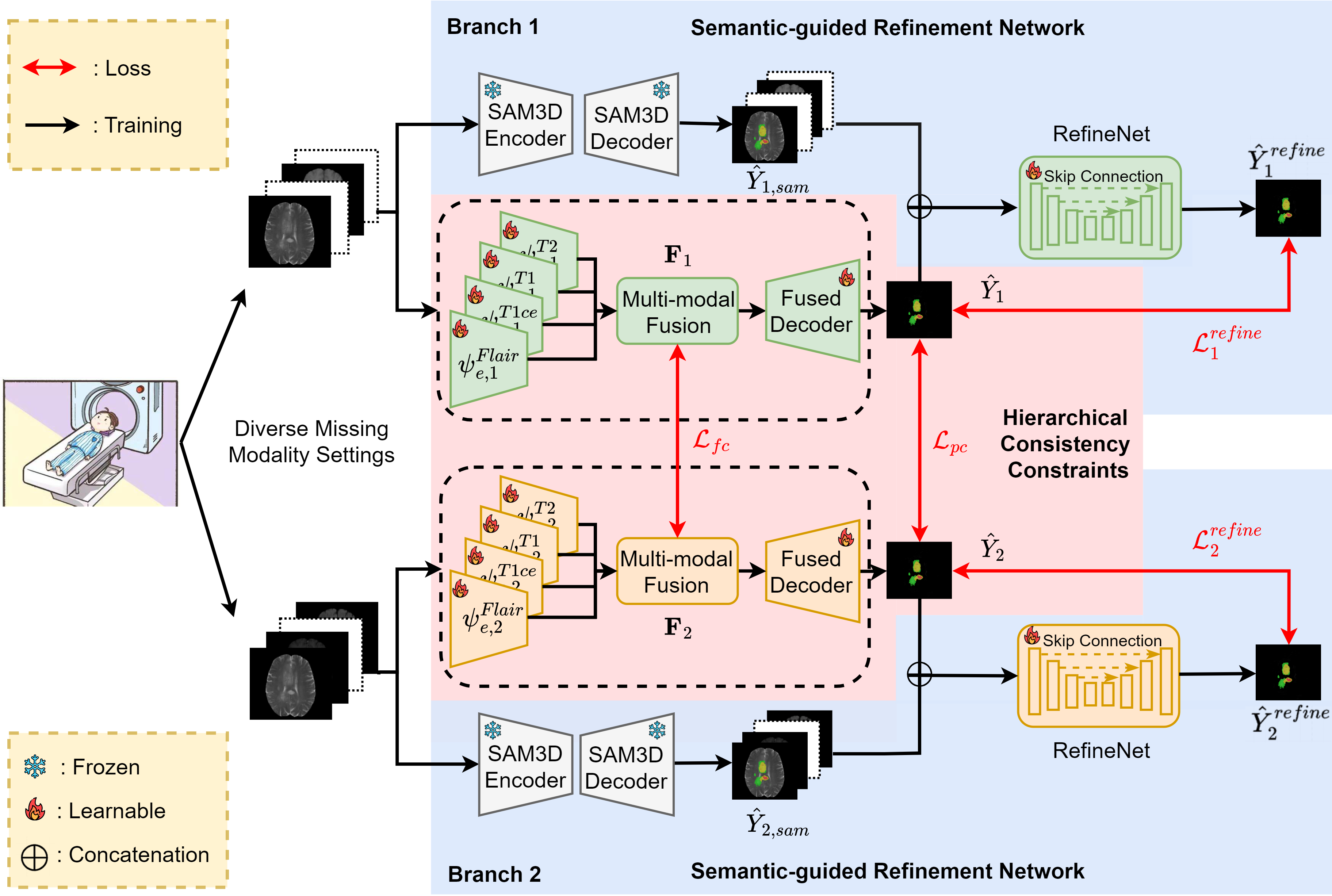}
\caption{Overview of the proposed Semantic-guided Masked Mutual Learning (SMML) framework.}
\label{fig1:env}
\end{figure*}
To improve the robustness and discriminative power of incomplete multi-modal learning, we propose a novel Semantic-guided Masked Mutual Learning (SMML) framework. Given the full-modal inputs, we generate random modality dropout masks for each student branch and feed the resulting missing modality data into each branch to foster the learning of robust multi-modal knowledge. An overview of the proposed framework is depicted in Figure \ref{fig1:env}, which consists of two student branches. With diverse missing modality inputs, the dual-branch networks are supervised by Hierarchical Consistency Constraints (HCC) in addition to the classic segmentation loss. During training, each student branch is further guided by a Semantic-guided Refinement Network (SRN) to enhance the discriminative power of each branch. In SRN, semantic priors derived from SAM are combined with the initial predictions from each student branch to generate refined outputs, which are then used to supervise the training of each branch.

\subsection{Semantic-guided Masked Mutual Learning }
\label{sec:MML}
Assuming patients undergoing varied missing modality imaging protocols should yield consistent outcomes, we propose the Semantic-guided Masked Mutual Learning (SMML) framework with two parallel branches. Denote the number of modalities (Flair, T1ce, T1, and T2) as $K$, for each branch $i, i \in \{1,2\}$, we introduce the modality-specific encoders $\psi_{e,i}^{k}(\cdot),i \in \{1,2\}, k \in \{1,\cdots, K\}$ to generate intermediate feature maps $\mathbf{F}^k_i \in \mathbb{R}^{H_i \times W_i \times Z_i \times d}$
for each modality input, where $(H_i, W_i, Z_i)$ and $d$ denote the spatial and channel dimensions of the intermediate feature maps.

For each branch $i$, given the complete multi-modal intermediate feature maps $\{\mathbf{F}^k_i, k \in \{1,\cdots, K\} \}$, we obtain the simulated missing modality inputs by randomly masking out one or more modality features. 
Denote the random modality dropout mask for each branch as $M_i = \{M_i^k\}$, where each element is a binary value in $\{0,1\}$, indicating the presence or absence of modality $k$ in branch $i$. And $M_i^k = 0$ when the $k$-th modality of the $i$-th branch is missing, otherwise $M_i^k = 1$. Then the features of the missing modality $k$ are substituted with zeros, while the features of the remaining modalities are left unchanged (i.e., $\mathbf{F}^k_i = \mathbf{F}^k_i * M_i^k$). The masked multi-modal feature set $\{\mathbf{F}^k_i,k \in 1,\cdots, K\}$ are further processed by an attention-based multi-modal feature fusion module~\cite{zhang2022mmformer} to generate the fused features $\mathbf{F}_1,\mathbf{F}_2 \in \mathbb{R}^{H_i \times W_i \times Z_i \times d_f}$ for each branch, where $d_f$ is the fused feature dimension.

 We then incorporate a fused decoder $\psi_{d,i}(\cdot)$ to generate the initial prediction for each branch as $\hat{Y}_i = \psi_{d,i}(\mathbf{F}_i ) $ . 
Each branch is further associated with a refinement network $f^{refine}_i(\cdot)$ to generate the refined segmentation predictions $\hat{Y}_i^{refine}$. The entire SMML framework is then supervised by the proposed Hierarchical Consistency Constraints and the Semantic-guided Refinement Network to enhance the robustness and discriminative ability of incomplete multi-modal learning with arbitrary missing modalities.

\subsection{Hierarchical Consistency Constraints (HCC)}
To leverage both detailed and relational context knowledge for robust learning, our HCC strategy involves consistency at both the pixel-level and the feature-level.
\paragraph{Pixel-level Bidirectional Constraint (PBC)}
Considering the sensitivity of different modalities to diverse tumor regions, the performance of each branch may suffer from severe performance degradation due to the absence of critical modality information. To improve the robustness of learning in the proposed SMML framework, we introduce pixel-level bidirectional distillation to selectively exchange reliable knowledge across the two branches.

With the initial predictions $\hat{Y}_1, \hat{Y}_2 \in \mathbb{R}^{H \times W \times Z \times C}$ from the two branches, where $H,W,Z$ correspond to the spatial dimensions of the input image, and $C$ is the number of classes. We proceed to compute the cross entropy loss between $\hat{Y}_1, \hat{Y}_2$ and the ground truth one hot segmentation map $Y_{oh}$ as,
\begin{equation}
\begin{aligned}
&Q_1 = \sum\nolimits_{c=1}^C- Y_{oh} \log(\sigma(\hat{Y}_1)), \\
&Q_2 = \sum\nolimits_{c=1}^C- Y_{oh} \log(\sigma(\hat{Y}_2)),
\end{aligned}
\end{equation}
where $\sigma(\cdot)$ is the softmax operation. The pixel-wise cross entropy loss $Q_1,Q_2  \in \mathbb{R}^{H \times W \times Z} $ for each branch measure the deviation from the ground truth segmentation map. A lower loss value indicates a prediction that is more reliable. Therefore, we can obtain the bidirectional transfer mask $T_M \in \mathbb{R}^{H \times W \times Z}$ as,
\begin{equation}
    T_M(h,w,z) = \mathds{1}[Q_1(h,w,z) > Q_2(h,w,z) ] ,
\end{equation}
where $\mathds{1}[\cdot]$ is the indicator function that evaluates to 1 when the inside condition is true, and 0 otherwise. $T_M(h,w,z)$ refers to the transferring direction at pixel $(h,w,z)$, and $T_M(h,w,z) =1$ means the reliable knowledge is exchanged from branch 2 to branch 1.

The pixel-wise consistency loss for each branch is then calculated based on the transferring mask as,
\begin{equation}
\begin{aligned}
    &\mathcal{L}_1^{pc} = \frac{1}{N_{21}}\sum\nolimits_{h,w,z} T_M \ast KL(\hat{Y}_{1} / \tau ||\hat{Y}_{2} / \tau ), \\
    &\mathcal{L}_2^{pc} = \frac{1}{N_{12}} \sum\nolimits_{h,w,z} (1-T_M) \ast KL(\hat{Y}_2 / \tau ||\hat{Y}_1 /\tau),
\end{aligned}
\end{equation}
where 
$N_{21} = \sum\nolimits_{h,w,z=1}^{H,W,Z} T_M(h,w,z)$, $N_{12} = H *W * Z - N_{21}$. $N_{21}$ denotes the number of pixels that facilitate the transfer of knowledge from branch 2 to branch 1, and vice versa for $N_{12}$. $\tau$ is the temperature hyper-parameter.
The process ensures that knowledge is reliably exchanged across the students, improving the segmentation performance.

\begin{figure*}[t]
\centering
\includegraphics[width=0.83\textwidth]{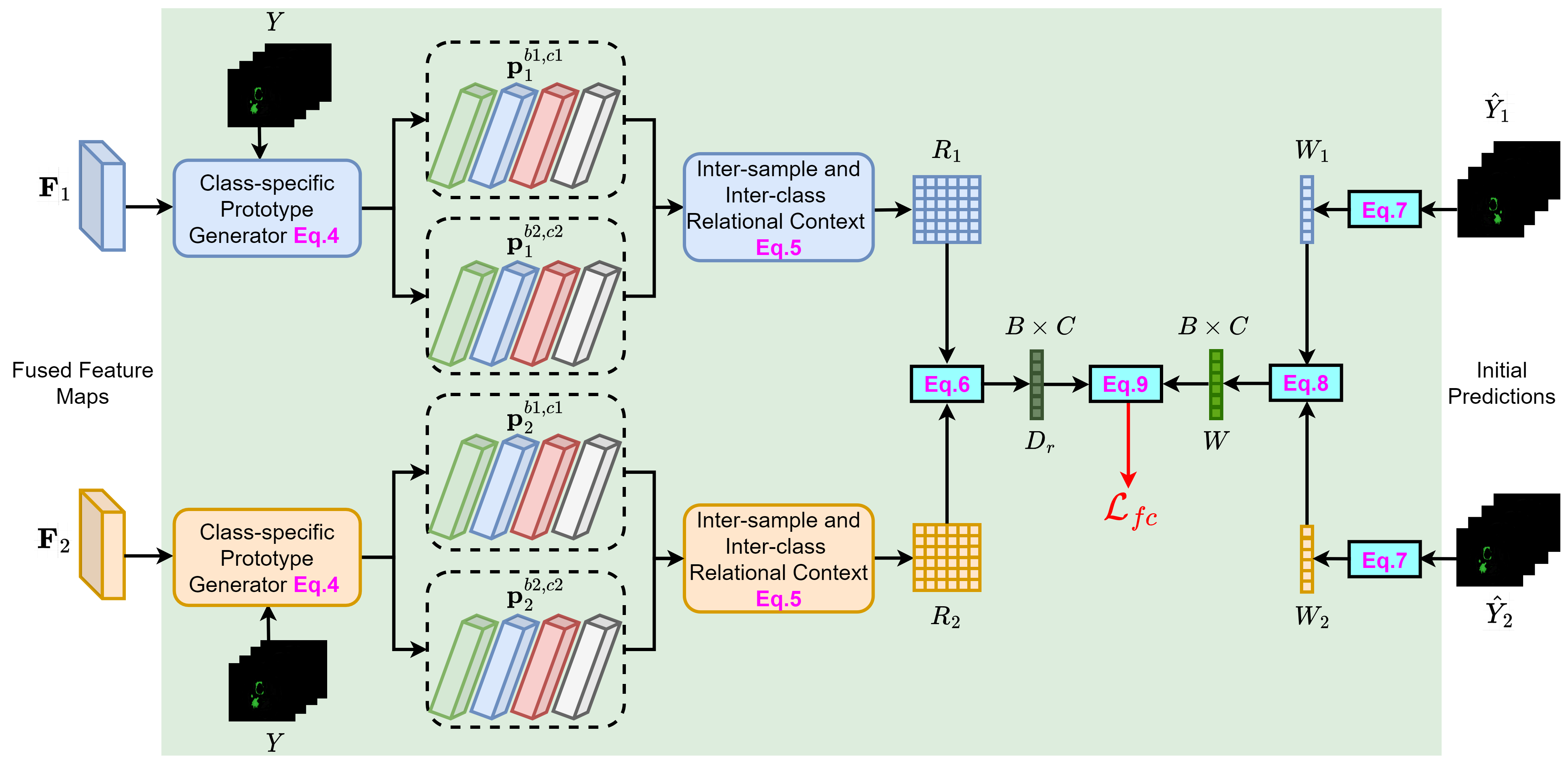}
\caption{Illustration of the Feature-level Relational Constraint (FRC) module. }
\label{fig2:env}
\end{figure*}

\paragraph{Feature-level Relational Constraint (FRC)}
Given that different branches should exhibit similar feature distributions, leveraging the feature affinity between inter-class prototypes (centers) is an effective strategy for robust relational context distillation in our masked mutual learning. As depicted in Figure \ref{fig2:env}, the FRC module comprises three key components: 1) a class-specific prototype generator, 2) a mechanism for generating inter-sample and inter-class relational context, and 3) an uncertainty-based re-weighting scheme to prioritize consistency learning from hard classes.

\textbf{class-specific prototype generator:} 
Given the fused features $\mathbf{F}_1,\mathbf{F}_2 \in \mathbb{R}^{H_i \times W_i \times Z_i \times d_f}$,
and the corresponding segmentation mask $Y \in \mathbb{R}^{H \times W \times Z}$, we first interpolate $Y$ into the same spatial dimension as $\mathbf{F}_1,\mathbf{F}_2 $, then we can get the prototype $\mathbf{p}_1^c, \mathbf{p}_2^c$ for each semantic class $c$ as,
\begin{equation}
\begin{aligned}
   & \mathbf{p}_1^c = \frac{1}{N_c}\sum\limits_{h,w,z=1}^{H_i,W_i,Z_i}\mathds{1}[Y(h,w,z) = c] * \mathbf{F}_1(h,w,z),\\
   & \mathbf{p}_2^c = \frac{1}{N_c}\sum\limits_{h,w,z=1}^{H_i,W_i,Z_i}\mathds{1}[Y(h,w,z) = c] * \mathbf{F}_2(h,w,z),
    \label{eq:proto2}
\end{aligned}
\end{equation}
where $N_c$ denotes the number of pixels belonging to class $c$.

\textbf{inter-sample and inter-class relational context modeling:} During the training of each mini-batch, denote the batchsize as $B$, the number of semantic classes as $C$, then according to Eq.~\ref{eq:proto2}, we can obtain the prototype of the $c$-th class in the $b$-th sample for each branch as $\mathbf{p}_1^{b,c} $ and $\mathbf{p}_2^{b,c}$, respectively. The inter-sample and inter-class relational context $R_1,R_2 \in  \mathbb{R}^{(B*C) \times (B*C)}$ for each branch can be calculated as,
\begin{equation}
\begin{aligned}
    & R_1(b_1,c_1;b_2,c_2) = \frac{<\mathbf{p}_1^{b_1,c_1},\mathbf{p}_1^{b_2,c_2}>}{||\mathbf{p}_1^{b_1,c_1}||_2||\mathbf{p}_1^{b_2,c_2}||_2}, \\
   & R_2(b_1,c_1;b_2,c_2) = \frac{<\mathbf{p}_2^{b_1,c_1},\mathbf{p}_2^{b_2,c_2}>}{||\mathbf{p}_2^{b_1,c_1}||_2||\mathbf{p}_2^{b_2,c_2}||_2},
\end{aligned}
\end{equation}

where $<,>$ calculates the dot product between two items, and $R_i(b_1,c_1;b_2,c_2)$ denotes the feature similarity between the $c_1$-th class of the $b_1$-th sample and the $c_2$-th class of the $b_2$-th sample in branch $i$. The initial inter-sample and inter-class relational distance is calculated as,
\begin{equation}
    D_r^{b_1,c_1 } = \sum_{b_2,c_2=1}^{B,C}(R_1(b_1,c_1;b_2,c_2) -R_2(b_1,c_1;b_2,c_2) )^2,
\end{equation}
where $D_r^{b_1,c_1}$ is the overall relational difference between the $c_1$-th class of the $b_1$-th sample across the two branches.

\textbf{uncertainty based re-weighting scheme:} 
To prioritize consistency learning on more challenging classes, we introduce a re-weighting scheme based on class-specific uncertainty. With the initial predictions $\hat{Y}_1, \hat{Y}_2 \in \mathbb{R}^{H \times W \times Z \times C}$, we first get the normalized prediction scores using softmax as $\sigma(\hat{Y}_i),i\in \{1,2\}$. Denote the normalized prediction scores of the $c$-th class as $\hat{Y}_i^c \in \mathbb{R}^{H \times W \times Z}$, 
then the class-specific weights $W_i = \{W_i^c\} \in \mathbb{R}^{C}$ are obtained as,
\begin{equation}
\begin{aligned}
   & W_1^c= \sum\nolimits -\hat{Y}_1^c log(\hat{Y}_1^c),\\
   &W_2^c = \sum\nolimits -\hat{Y}_2^c log(\hat{Y}_2^c).
\end{aligned}
\end{equation}

The final uncertainty weights are the mean of weights from both branches, 
\begin{equation}
   W^c = ( W_1^c + W_2^c)/2.
\end{equation}
Similarly, we can calculate the class-specific weights for each sample within the mini-batch. Denote $W^{b_1,c_1}$ as the weight for the $c_1$-th class of the $b_1$-th sample, the final feature-level relational consistency loss is calculated as,
\begin{equation}
    \mathcal{L}_{fc} =\sum\nolimits_{b_1,c_1=1}^{B,C} D_r^{b_1,c_1 } * W^{b_1,c_1}.
\end{equation}

\subsection{Semantic-guided Refinement Network}

To further enhance the discriminative power of our method, we propose incorporating auxiliary semantic cues from the well-established SAM model \cite{kirillov2023segment}. While the SAM model has demonstrated impressive zero-shot segmentation capabilities in 2D natural images, its efficacy is significantly compromised when applied to medical images, primarily due to the lack of medical domain expertise. Consequently, we opt for the SAM-Med3D model \cite{wang2023sammed3d}, which has been specifically trained on a variety of 3D medical imaging tasks, to produce supplementary semantic segmentation cues. 

For each branch $i$, with the initial segmentation predictions $\hat{Y}_i$, and the zero-shot segmentation results $\hat{Y}_{i,sam}^k$ generated from each modality $k$ using SAM-Med3D, we first substitute the missing modality results with all zeros (i.e., $\hat{Y}_{i,sam}^k =0 | M_i^k = 0$, where $M_i^k$ refers to the previously defined random modality dropout mask), then the multi-modal semantic cues from SAM, denoted as $\hat{Y}_{i,sam} = \{\hat{Y}_{i,sam}^k, k \in 1,\cdots, K\}$, are concatenated with the initial prediction along the channel dimension. The combined predictions are further fed forward through a refinement network to get the refined segmentation results,
\begin{equation}
    \hat{Y}_i^{refine} = f^{refine}_i(Cat[\hat{Y}_{i,sam},\hat{Y}_i]),
\end{equation}
where $f^{refine}_i(\cdot)$ is the refinement network of branch $i$, which is a simple UNet model, $Cat[,]$ refers to concatenation along the channel dimension. The refined segmentation results are then utilized to guide the training by measuring the consistency between the initial and refined predictions,
\begin{equation}
\begin{aligned}
   & L_{1}^{refine} = KL(\hat{Y}_1|| \hat{Y}_1^{refine}) ,\\
   & L_{2}^{refine} = KL(\hat{Y}_2|| \hat{Y}_2^{refine}) .
    \end{aligned}
\end{equation}
\textit{Please be aware that the refinement network operates solely during the training phase and does not impose any computational overhead during inference.}

\subsection{Training and Inference}
Denote the classic segmentation loss on the initial predictions $\hat{Y}_i$ and the refined predictions $\hat{Y}_i^{refine}$ as,
\begin{equation}
\begin{aligned}
    L_1^{task} &= CE(\hat{Y}_1,Y)+Dice(\hat{Y}_1,Y),\\
    L_2^{task} &= CE(\hat{Y}_2,Y)+Dice(\hat{Y}_2,Y),\\
    L_{1,refine}^{task} &= CE(\hat{Y}_1^{refine},Y)+Dice(\hat{Y}_1^{refine},Y),\\
    L_{2,refine}^{task} &= CE(\hat{Y}_2^{refine},Y)+Dice(\hat{Y}_2^{refine},Y),
\end{aligned}
\end{equation}
where $CE(\cdot),Dice(\cdot)$ refer to the cross entropy and dice losses, respectively. Then the overall objectives of the proposed method for each branch is formulated as,
\begin{equation}
\begin{aligned}
    &\mathcal{L}_1 = \mathcal{L}^{task}_1 + \mathcal{L}^{task}_{1, refine}  + \mathcal{L}^{pc}_1 +  \mathcal{L}_{fc}+ \mathcal{L}_1^{refine}, \\
    &\mathcal{L}_2 = \mathcal{L}^{task}_2 + \mathcal{L}^{task}_{2, refine} + \mathcal{L}^{pc}_2 + \mathcal{L}_{fc} + 
    \mathcal{L}_2^{refine}.
\end{aligned}
\end{equation}

The final loss objective for each branch, denoted as $L_i, i \in \{1,2\}$, is used to guide the learning process of the respective branch during training.

During inference, to assess performance under each missing modality setting, we duplicate inputs with the arbitrary missing modalities and feed them to each branch to generate the initial predictions $\hat{Y}_1$ and $\hat{Y}_2$. The final segmentation mask is predicted by taking the mean of $\hat{Y}_1$ and $\hat{Y}_2$.

\begin{table*}[ht]
    \centering
    \scalebox{0.83}{
    \begin{tabular}{|c|c|c|c|c|c|c|c|c|c|c|c|c|c|c|c|c|c|}
        \toprule
    \multirow{4}{*}{M}&Flair&$\bullet$&$\circ$  &$\circ$  &$\circ$    &$\bullet$&$\circ$  &$\circ$  &$\bullet$&$\circ$  &$\bullet$   &$\circ$  &$\bullet$&$\bullet$&$\bullet$  &$\bullet$&\multirow{4}{*}{Avg} \\
                      &T1ce &$\circ$  &$\bullet$&$\circ$  &$\circ$    &$\bullet$&$\circ$  &$\bullet$&$\circ$  &$\bullet$&$\circ$     &$\bullet$&$\bullet$&$\circ$  &$\bullet$  &$\bullet$&    \\
                      &T1   &$\circ$  &$\circ$  &$\bullet$&$\circ$    &$\circ$  &$\bullet$&$\bullet$&$\circ$  &$\circ$  &$\bullet$   &$\bullet$&$\circ$  &$\bullet$&$\bullet$  &$\bullet$&   \\
                      &T2   &$\circ$  &$\circ$  &$\circ$  &$\bullet$  &$\circ$  &$\bullet$&$\circ$  &$\bullet$&$\bullet$&$\circ$     &$\bullet$&$\bullet$&$\bullet$&$\circ$    &$\bullet$&    \\
        \midrule
    \multirow{7}{*}{WT}&MCTSeg&84.3&73.8&73.1&84.4&88.6&86.0&79.2&88.5&86.5&88.1&87.0&89.4&89.0&88.8&89.6&85.1 \\
                      &M3AE&\textbf{88.7}&75.8&74.4&84.8&\textcolor{blue}{89.7}&86.7&77.2&89.9&86.3&89.0&85.7&\textcolor{blue}{90.2}&\textcolor{blue}{89.9}&88.9&\textcolor{blue}{90.6}&85.8  \\
                      &U-Net-MFI&86.4&75.2&74.2&85.6&\textbf{90.0}&86.6&78.7&\textbf{90.0}&86.5&\textcolor{blue}{89.5}&86.9&\textbf{90.8}&\textbf{90.3}&\textcolor{blue}{90.1}&\textbf{90.9}&\textcolor{blue}{86.1} \\                   
                     &RFNet&84.4&74.9&73.6&84.8&88.4&86.6&78.4&88.6&86.8&87.8&87.2&89.5&89.2&88.9&89.6&85.3   \\&mmFormer&85.7&77.9&76.5&\textcolor{blue}{86.0}&88.7&\textcolor{blue}{86.8}&\textcolor{blue}{80.4}&88.7&\textcolor{blue}{86.9}&88.1&\textcolor{blue}{87.1}&89.7&89.1&89.0&89.7&86.0   \\
                      &SMU-Net&87.5&\textbf{80.3}&\textcolor{blue}{78.6}&85.7&88.4&85.6&80.3&87.9&86.1&87.3&86.5&88.2&88.3&88.2&88.9&85.9                \\
                        &Ours&\textcolor{blue}{87.9}&\textcolor{blue}{79.8}&\textbf{78.7}&\textbf{87.0}&89.5&\textbf{87.6}&\textbf{82.8}&\textcolor{blue}{89.9}&\textbf{87.6}&\textbf{89.7}&\textbf{87.9}&89.9&89.7&\textbf{90.3}&89.9&\textbf{87.2} 
  \\
        \midrule
    \multirow{7}{*}{TC}&MCTSeg&59.9&79.7&56.4&66.2&82.4&68.6&82.3&69.5&83.4&67.4&83.6&83.2&70.3&82.9&83.0&74.6 \\
            &M3AE&66.4&82.9&\textcolor{blue}{66.1}&69.4&84.4&71.8&83.4&70.9&84.2&70.8&84.4&84.6&72.7&84.1&84.5&77.4 \\  
            &U-Net-MFI&63.7&80.9&63.1&67.5&\textcolor{blue}{84.7}&71.4&\textbf{91.9}&71.3&84.0&\textcolor{blue}{72.7}&84.1&\textcolor{blue}{85.5}&\textcolor{blue}{74.0}&\textbf{85.2}&\textcolor{blue}{85.6}&77.0 \\
                     &RFNet&63.5&79.9&61.1&65.7&81.9&71.1&81.2&70.8&83.1&71.5&83.9&83.4&72.8&83.2&84.0&75.8  \\&mmFormer&62.9&81.9&62.1&\textcolor{blue}{69.7}&82.7&69.6&82.8&\textcolor{blue}{71.3}&83.4&68.7&83.9&83.3&70.6&83.5&83.9&76.0 \\ 
                      &SMU-Net&\textbf{71.8}&\textbf{84.1}&\textbf{69.5}&67.2&84.1&\textcolor{blue}{73.5}& \textcolor{blue}{84.4}&71.2&\textbf{85.0}&71.2&\textcolor{blue}{84.4}&82.5&67.9&84.2&\textbf{87.3}&\textcolor{blue}{77.9}       \\
                      &Ours&\textcolor{blue}{67.7}&\textcolor{blue}{83.1}&65.4&\textbf{71.7}&\textbf{84.8}&\textbf{73.6}&84.0&\textbf{73.4}&\textcolor{blue}{84.8}&\textbf{72.9}&\textbf{84.9}&\textbf{86.3}&\textbf{74.9}&\textcolor{blue}{84.7}&85.2&\textbf{78.5}  \\
        \midrule
    \multirow{7}{*}{ET}&MCTSeg&37.7&73.3&28.1&37.2&74.8&40.1&73.1&43.1&76.9&39.7&75.3&75.3&\textcolor{blue}{46.5}&75.4&75.4&58.1  \\
                        &M3AE&35.6&73.7&37.1&\textcolor{blue}{47.6}&75.0&\textbf{48.7}&74.7&45.4&75.3&41.2&75.4&73.8&44.8&74.0&75.5&59.9 \\
                   &U-Net-MFI&37.6&76.2&32.3&37.4&\textcolor{blue}{78.8}&41.4&76.8&42.5&77.6&41.3&77.9&\textcolor{blue}{79.4}&43.4&\textcolor{blue}{79.8}&\textcolor{blue}{80.0}&60.2 \\
                    
                     &RFNet&34.7&70.0&30.2&40.8&71.6&42.7&72.8&44.9&72.3&39.5&74.5&73.5&46.1&74.1&74.3&57.5  \\&mmFormer&39.0&75.3&37.7&45.8&76.0&45.0&\textcolor{blue}{78.1}&44.5&\textcolor{blue}{77.8}&43.2&\textcolor{blue}{78.5}&77.6&46.3&79.2&79.4&61.5 \\
                      &SMU-Net&\textbf{46.1}&\textcolor{blue}{78.3}&\textbf{42.8}&43.1&77.1&\textcolor{blue}{47.7}&75.1&\textcolor{blue}{46.0}&75.7&\textcolor{blue}{44.0}&76.2&75.4&43.1&76.2&79.3&\textcolor{blue}{61.8}         \\
                &Ours&\textcolor{blue}{41.0}&\textbf{79.1}&\textcolor{blue}{39.4}&\textbf{47.8}&\textbf{81.8}&45.2&\textbf{79.9}&\textbf{46.5}&\textbf{81.6}&\textbf{45.1}&\textbf{81.6}&\textbf{81.5}&\textbf{48.9}&\textbf{81.5}&\textbf{81.7}&\textbf{64.2}  \\

        \bottomrule
    \end{tabular}}
    \caption{Comparisons with the SOTA methods on the BraTs 2018 dataset.  Performance is evaluated based on Dice Similarity Coefficient (DSC) scores for the whole tumor (WT), tumor core (TC), and enhancing tumor (ET). Hollow circles indicate the absence of that modality. The best results are highlighted in \textbf{bold}, while the second-best scores are indicated in \textcolor{blue}{blue}.}
    \label{tab1:env}
\end{table*}
\section{Experiments and Results}
We demonstrate the effectiveness of our SMML framework through comprehensive comparison with state-of-the-art methods and an ablation study in the subsequent sections. Additionally, we validate its generalization capability with experiments on incomplete multi-modal segmentation of natural images, as detailed in the \textcolor{magenta}{supplementary material}.

\subsection{Datasets and Evaluation Metrics}
The subjects in the BraTs 2018 and BraTs 2020 datasets \cite{menze2014multimodal} are categorized into four subregions: background, necrosis, edema, and enhancing tumor. In contrast, images in BraTs 2015 are divided into five subregions, which include the aforementioned categories plus an additional one for non-enhancing tumor.
These subregions are further consolidated into three nested subregions: whole tumor, tumor core, and enhancing tumor. 
All volumetric data have been co-registered to a standard anatomical template and interpolated to a uniform resolution as provided by the dataset organizers.

Consistent with prior approaches \cite{ding2021rfnet,liu2023m3ae}, 
the BraTs 2015 dataset, comprising 274 subjects with public ground truth, is distributed into training, validation, and testing sets of
242, 12, and 20 samples, respectively. In the BraTs 2018 dataset, 285 annotated samples are allocated into training, validation, and testing sets of 199, 29, and 57, respectively. Likewise, BraTs 2020's labeled images are split into training, validation, and testing sets of 219, 50, and 100, respectively.
The Dice Similarity Coefficient (DSC) is employed as the evaluation metric to assess the performance of brain tumor segmentation.

\subsection{Implementation Details}

All experiments were conducted using Python 3.9, PyTorch 2.1.1, and Ubuntu 22.04. The proposed SMML framework was trained on two NVIDIA A800 GPUs with 80GB of memory. The Adam optimizer was used for model optimization, with $\beta_1$ and $\beta_2$ parameters set to 0.9 and 0.999 respectively, and a weight decay of 1e-4. The ‘poly' learning rate policy was adopted, starting with an initial learning rate of 2e-4. 
The temperature hyper-parameter $\tau$ is empirically set to 6. 
During training, input images were randomly cropped to $128 \times 128 \times 128$ and then augmented using random flipping, cropping, and intensity shifts. The model was trained for 1000 epochs with a batch size of 2. 

\begin{figure*}[ht]
\centering
\includegraphics[width=0.9\textwidth]{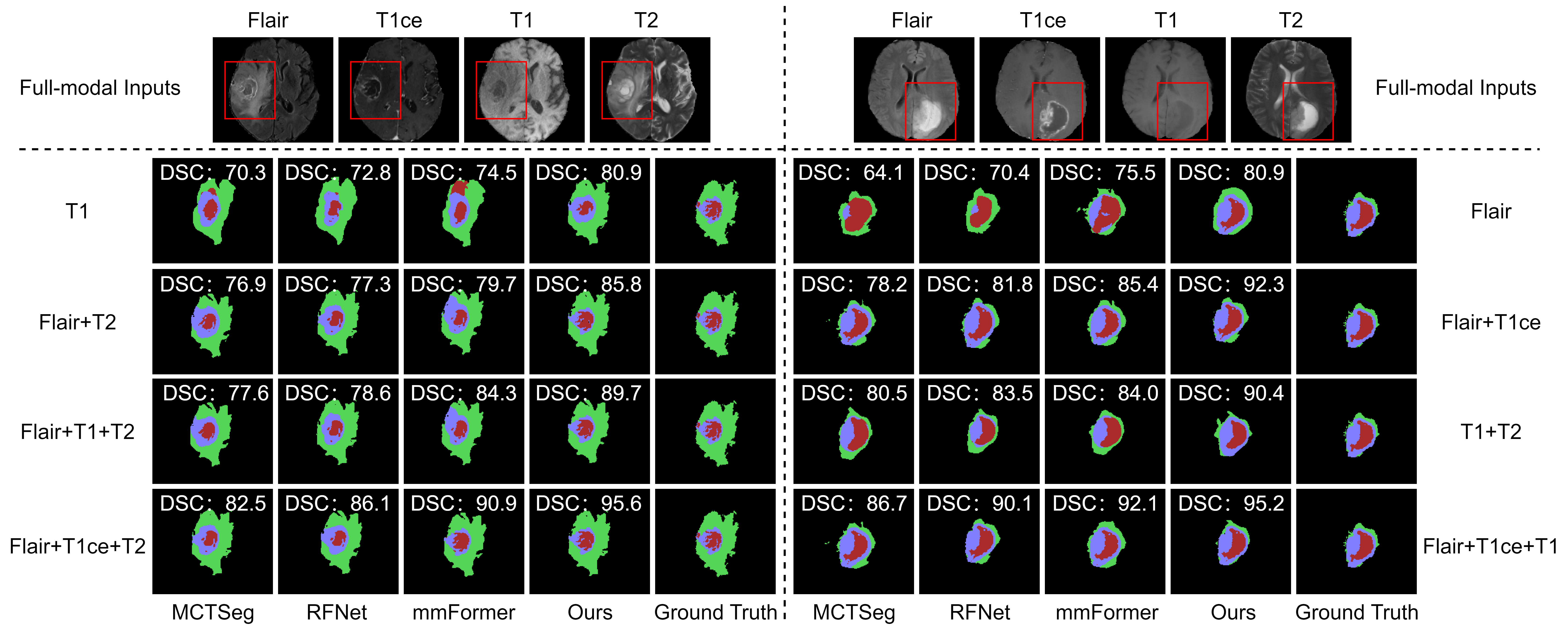}
\caption{ The first row shows full-modal inputs for two samples. Below are predicted segmentation masks for various missing modality settings. Green: peritumoral edema; blue: enhancing tumor; red: necrotic and non-enhancing tumor core.}
\label{fig3:env}
\end{figure*}

\begin{table}[ht]
    \centering
    \scalebox{0.88}{
    \begin{tabular}{c|c|c|c|c}
        \toprule
        \multirow{2}{*}{Methods}&\multicolumn{4}{c}{Avg DSC scores (\%)}     \\ 
        \cline{ 2-5 } 
        & WT& TC& ET&Mean   \\
        \midrule
        MCTSeg       &85.8&66.6&60.8&71.1   \\
        M3AE         &85.6&67.6&59.7&71.0   \\   
        U-Net-MFI   &85.7& 69.3&61.2&72.1   \\
        RFNet        &85.4&71.1&63.0&73.2 \\     
        mmFormer     &86.9&71.7&62.4&73.7  \\ 
        SMU-Net      &86.7&72.1&62.7&73.8  \\
        \midrule
        Ours         &\textbf{87.6}&\textbf{72.8}&\textbf{64.9}&\textbf{75.1}  \\ 
        \bottomrule
    \end{tabular}}
    \caption{Comparison results on the BraTs 2015 dataset. }
    \label{tab2:env}
\end{table}

\subsection{Comparison with State-of-the-art Methods}
We compare the proposed SMML with SOTA methods, including RFNet \cite{ding2021rfnet}, mmFormer \cite{zhang2022mmformer}, U-Net-MFI \cite{zhao2022modality}, SMU-Net \cite{azad2022smu}, M3AE \cite{liu2023m3ae} and MCTSeg \cite{kang2024multimodal}.

We evaluate our method on three challenging multi-modal brain tumor segmentation datasets. 
We record the detailed results under different scenarios with arbitrary missing modality settings. The quantitative comparison results on the BraTs 2015, BraTs 2018, and BraTs 2020 datasets are presented in Table \ref{tab1:env}, Table \ref{tab2:env}, and Table \ref{tab3:env}, respectively. On BraTs 2018, we provide detailed comparison results for all missing modality settings. For BraTs 2015 and BraTs 2020, we report the average DSC scores across fifteen different missing modality combinations for brevity.  
MCTSeg, M3AE, U-Net-MFI, mmFormer, and SMU-Net were evaluated on only one or two datasets. For a thorough comparison, we replicated their results on the remaining datasets using the official implementations. 

As shown in Tables \ref{tab1:env}, \ref{tab2:env}, and \ref{tab3:env}, 
our approach markedly improves segmentation accuracy, achieving state-of-the-art performance across all three datasets and tumor categories.
Specifically, our SMML approach outperforms the second-best method with mean DSC score gains of 1.4\% on BraTs 2018, 1.3\% on BraTs 2015 and 1.7\% on BraTs 2020, respectively.
In the detailed comparison of scenarios with arbitrary missing modalities (as shown in Table~\ref{tab1:env}), our proposed SMML achieves the highest or runner-up DSC scores in the majority of missing modality configurations across the three semantic categories.

Figure \ref{fig3:env} displays qualitative comparison results, featuring two samples from the BraTs 2018 dataset.
The comparison demonstrates that our method accurately identifies the tumor locations and covers the majority of tumor regions in different modal combination scenarios, while previous methods exhibit incomplete or incorrect tumor delineation.
Furthermore, we note that previous methods exhibit suboptimal segmentation performance when only a single Flair or T1 image is available, due to the limited information from single modality.
In contrast, our SMML maintains satisfactory performance under these conditions, illustrating the strength of our semantic-guided masked mutual learning in capturing robust and discriminative features for segmentation.
\begin{table}[ht]
    \centering
    \scalebox{0.88}{
    \begin{tabular}{c|c|c|c|c}
        \toprule
        \multirow{2}{*}{Methods}&\multicolumn{4}{c}{Avg DSC scores (\%)}     \\   
        \cline{ 2-5 }
        & WT& TC& ET&Mean   \\
        \midrule
        MCTSeg       &85.4&75.1&60.5&73.7   \\ 
        M3AE         &86.9&79.1&61.7&75.9   \\
        U-Net-MFI    &86.1&78.9&61.9&75.6    \\
        RFNet        &86.3&78.0&61.4&75.2 \\
        mmFormer     &86.5&78.3&65.1&76.6  \\ 
        SMU-Net      &86.4&79.1&64.9&76.8 \\
        \midrule
        Ours         &\textbf{87.7}&\textbf{81.2}&\textbf{66.4}&\textbf{78.5} \\ 
        \bottomrule
    \end{tabular}}
    \caption{Comparison results on the BraTs 2020 dataset. }
    \label{tab3:env}
\end{table}

\subsection{Ablation Study}
We evaluate the impact of each novel component in our SMML through ablation studies, incrementally integrating them into the baseline mmFormer model on the BraTs 2018 dataset.
Table \ref{tab4:env} illustrates the average DSC scores across various missing modality scenarios for each category, with an additional overall mean for direct comparison. 
As shown, the semantic-guided refinement network (SRN) achieves a 0.7\% improvement over the baseline, confirming the benefit of auxiliary semantic information from SAM. Additionally, the dual-branch scheme alone surpasses the baseline by 0.8\% in DSC, underscoring the efficacy of masked mutual learning.
Based on the dual-branch scheme, 
the PBC, FRC, and SRN modules can further lead to notable improvements of 0.4\%, 0.5\% and 0.4\%, respectively.
Furthermore, the various combinations of different modules also lead to remarkable improvements.
These highlight the indispensable role of these modules in learning robust and discriminative representations for handling arbitrary missing modalities.
Ultimately, integrating all proposed modules, the SMML framework attains a mean DSC score of 76.7\% across the three categories, exceeding the baseline by 2.2\%.

\begin{table}[ht]
    \centering
    \scalebox{0.77}{
    \begin{tabular}{c|c|c|c|c|c|c|c}
        \toprule
        \multirow{2}{*}{Dual-branch}&\multirow{2}{*}{PBC}&\multirow{2}{*}{FRC}&\multirow{2}{*}{SRN}& \multicolumn{4}{c}{DSC scores (\%)}  \\ 
        \cline{5-8}
         &&&& WT& TC& ET&Mean   \\
         \midrule
        &&&                           &86.0&76.0&61.5&74.5 \\
        \midrule
       &&  & $\checkmark$             &86.2&77.6&61.9&75.2 \\
        $\checkmark$&               &&&86.3&77.3&62.4&75.3 \\
        $\checkmark$ & $\checkmark$ &&&86.8&78.0&62.3&75.7 \\
        $\checkmark$&& $\checkmark$  &&86.5&77.6&63.2&75.8  \\
        $\checkmark$& &&  $\checkmark$ &86.5&77.4&63.2&75.7 \\
         $\checkmark$ & $\checkmark$ &$\checkmark$&&86.6&78.1&63.4&76.1 \\
         $\checkmark$ & $\checkmark$ &&$\checkmark$&86.7&78.3&62.7&75.9 \\
         $\checkmark$ && $\checkmark$ &$\checkmark$&86.9&78.5&63.2&76.2 \\
        \midrule
        $\checkmark$ & $\checkmark$ & $\checkmark$ & $\checkmark$ &\textbf{87.2}&\textbf{78.5}&\textbf{64.2}&\textbf{76.7}  \\ 
        \bottomrule
    \end{tabular}}
    \caption{Effectiveness of each SMML component.
    }
    \label{tab4:env}
\end{table}

\section{Conclusion}
We propose a novel Semantic-guided Masked Mutual Learning (SMML) method for brain tumor segmentation that flexibly addresses the challenges posed by diverse missing modality situations. Our SMML approach introduces a masked mutual learning scheme to emulate patient scenarios with varying missing modalities. We integrate Hierarchical Consistency Constraints (HCC) to transfer reliable knowledge and capture inter-sample and inter-class relational context in the latent feature space, enabling robust training of the SMML framework.
Additionally, we incorporate a refinement network into each branch that leverages semantic priors from SAM, further enhancing the discriminative capability of our method without introducing additional computational burden during inference.
Extensive experiments demonstrate that our SMML method significantly improves brain tumor segmentation performance under various missing modality settings.

\section*{Acknowledgements}
This study was supported under the Key Research and Development Program of Zhejiang Province (Grant No. 2023C03192). 
It was also funded by the National Science Foundation of China (Grant No. 62201341).

\bibliography{aaai25}

\begin{thebibliography}{26}
\providecommand{\natexlab}[1]{#1}

\bibitem[{Azad, Khosravi, and Merhof(2022)}]{azad2022smu}
Azad, R.; Khosravi, N.; and Merhof, D. 2022.
\newblock SMU-Net: Style matching U-Net for brain tumor segmentation with missing modalities.
\newblock In \emph{International Conference on Medical Imaging with Deep Learning}, 48--62. PMLR.

\bibitem[{Bakas et~al.(2017)Bakas, Akbari, Sotiras, Bilello, Rozycki, Kirby, Freymann, Farahani, and Davatzikos}]{bakas2017advancing}
Bakas, S.; Akbari, H.; Sotiras, A.; Bilello, M.; Rozycki, M.; Kirby, J.~S.; Freymann, J.~B.; Farahani, K.; and Davatzikos, C. 2017.
\newblock Advancing the cancer genome atlas glioma MRI collections with expert segmentation labels and radiomic features.
\newblock \emph{Scientific data}, 4(1): 1--13.

\bibitem[{Ding, Yu, and Yang(2021)}]{ding2021rfnet}
Ding, Y.; Yu, X.; and Yang, Y. 2021.
\newblock RFNet: Region-aware fusion network for incomplete multi-modal brain tumor segmentation.
\newblock In \emph{Proceedings of the IEEE/CVF international conference on computer vision}, 3975--3984.

\bibitem[{Han et~al.(2020)Han, Zhang, Fu, and Zhou}]{han2020trusted}
Han, Z.; Zhang, C.; Fu, H.; and Zhou, J.~T. 2020.
\newblock Trusted multi-view classification.
\newblock In \emph{International Conference on Learning Representations}.

\bibitem[{Hinton, Vinyals, and Dean(2015)}]{hinton2015distilling}
Hinton, G.; Vinyals, O.; and Dean, J. 2015.
\newblock Distilling the Knowledge in a Neural Network.
\newblock \emph{Computer Science}, 14(7): 38--39.

\bibitem[{Huang and Wang(2017)}]{huang2017like}
Huang, Z.; and Wang, N. 2017.
\newblock Like what you like: Knowledge distill via neuron selectivity transfer.
\newblock \emph{arXiv preprint arXiv:1707.01219}.

\bibitem[{Kang et~al.(2024)Kang, Ting, Phan, Ge, and Ting}]{kang2024multimodal}
Kang, M.; Ting, F.~F.; Phan, R. C.-W.; Ge, Z.; and Ting, C.-M. 2024.
\newblock A Multimodal Feature Distillation with CNN-Transformer Network for Brain Tumor Segmentation with Incomplete Modalities.
\newblock \emph{arXiv preprint arXiv:2404.14019}.

\bibitem[{Karimijafarbigloo et~al.(2023)Karimijafarbigloo, Azad, Kazerouni, Ebadollahi, and Merhof}]{karimijafarbigloo2023mmcformer}
Karimijafarbigloo, S.; Azad, R.; Kazerouni, A.; Ebadollahi, S.; and Merhof, D. 2023.
\newblock {MMCF}ormer: Missing Modality Compensation Transformer for Brain Tumor Segmentation.
\newblock In \emph{Medical Imaging with Deep Learning}.

\bibitem[{Kirillov et~al.(2023)Kirillov, Mintun, Ravi, Mao, Rolland, Gustafson, Xiao, Whitehead, Berg, Lo et~al.}]{kirillov2023segment}
Kirillov, A.; Mintun, E.; Ravi, N.; Mao, H.; Rolland, C.; Gustafson, L.; Xiao, T.; Whitehead, S.; Berg, A.~C.; Lo, W.-Y.; et~al. 2023.
\newblock Segment anything.
\newblock In \emph{Proceedings of the IEEE/CVF International Conference on Computer Vision}, 4015--4026.

\bibitem[{Li et~al.(2021)Li, Lei, Sun, and Kuang}]{li2021dynamic}
Li, X.; Lei, L.; Sun, Y.; and Kuang, G. 2021.
\newblock Dynamic-hierarchical attention distillation with synergetic instance selection for land cover classification using missing heterogeneity images.
\newblock \emph{IEEE Transactions on Geoscience and Remote Sensing}, 60: 1--16.

\bibitem[{Liu et~al.(2021)Liu, Tan, Wan, Liang, Lei, Guo, and Li}]{liu2021face}
Liu, A.; Tan, Z.; Wan, J.; Liang, Y.; Lei, Z.; Guo, G.; and Li, S.~Z. 2021.
\newblock Face anti-spoofing via adversarial cross-modality translation.
\newblock \emph{IEEE Transactions on Information Forensics and Security}, 16: 2759--2772.

\bibitem[{Liu et~al.(2023)Liu, Wei, Lu, Sun, Wang, and Zheng}]{liu2023m3ae}
Liu, H.; Wei, D.; Lu, D.; Sun, J.; Wang, L.; and Zheng, Y. 2023.
\newblock M3AE: multimodal representation learning for brain tumor segmentation with missing modalities.
\newblock In \emph{Proceedings of the AAAI Conference on Artificial Intelligence}, 2, 1657--1665.

\bibitem[{Menze et~al.(2014)Menze, Jakab, Bauer, Kalpathy-Cramer, Farahani, Kirby, Burren, Porz, Slotboom, Wiest et~al.}]{menze2014multimodal}
Menze, B.~H.; Jakab, A.; Bauer, S.; Kalpathy-Cramer, J.; Farahani, K.; Kirby, J.; Burren, Y.; Porz, N.; Slotboom, J.; Wiest, R.; et~al. 2014.
\newblock The multimodal brain tumor image segmentation benchmark (BRATS).
\newblock \emph{IEEE transactions on medical imaging}, 34(10): 1993--2024.

\bibitem[{Qiu et~al.(2023)Qiu, Chen, Yao, Xu, and Wang}]{qiu2023scratch}
Qiu, Y.; Chen, D.; Yao, H.; Xu, Y.; and Wang, Z. 2023.
\newblock Scratch Each Other's Back: Incomplete Multi-Modal Brain Tumor Segmentation via Category Aware Group Self-Support Learning.
\newblock In \emph{Proceedings of the IEEE/CVF International Conference on Computer Vision}, 21317--21326.

\bibitem[{Shaker et~al.(2024)Shaker, Maaz, Rasheed, Khan, Yang, and Khan}]{10526382unetr}
Shaker, A.~M.; Maaz, M.; Rasheed, H.; Khan, S.; Yang, M.-H.; and Khan, F.~S. 2024.
\newblock UNETR++: Delving into Efficient and Accurate 3D Medical Image Segmentation.
\newblock \emph{IEEE Transactions on Medical Imaging}, 1--1.

\bibitem[{Tung and Mori(2019)}]{tung2019similarity}
Tung, F.; and Mori, G. 2019.
\newblock Similarity-preserving knowledge distillation.
\newblock In \emph{Proceedings of the IEEE/CVF international conference on computer vision}, 1365--1374.

\bibitem[{Wang et~al.(2023)Wang, Guo, Ye, Deng, Cheng, Li, Chen, Su, Huang, Shen, Fu, Zhang, He, and Qiao}]{wang2023sammed3d}
Wang, H.; Guo, S.; Ye, J.; Deng, Z.; Cheng, J.; Li, T.; Chen, J.; Su, Y.; Huang, Z.; Shen, Y.; Fu, B.; Zhang, S.; He, J.; and Qiao, Y. 2023.
\newblock SAM-Med3D.
\newblock arXiv:2310.15161.

\bibitem[{Wang et~al.(2022)Wang, Zhang, Chen, Ma, Avery, Hull, and Carneiro}]{wang2022uncertainty}
Wang, H.; Zhang, J.; Chen, Y.; Ma, C.; Avery, J.; Hull, L.; and Carneiro, G. 2022.
\newblock Uncertainty-aware multi-modal learning via cross-modal random network prediction.
\newblock In \emph{European Conference on Computer Vision}, 200--217. Springer.

\bibitem[{Wei, Luo, and Luo(2023)}]{wei2023mmanet}
Wei, S.; Luo, C.; and Luo, Y. 2023.
\newblock MMANet: Margin-aware distillation and modality-aware regularization for incomplete multimodal learning.
\newblock In \emph{Proceedings of the IEEE/CVF Conference on Computer Vision and Pattern Recognition}, 20039--20049.

\bibitem[{Wu et~al.(2023)Wu, Lin, Huang, Fan, and Li}]{wu2023extracting}
Wu, L.; Lin, H.; Huang, Y.; Fan, T.; and Li, S.~Z. 2023.
\newblock Extracting low-/high-frequency knowledge from graph neural networks and injecting it into mlps: An effective gnn-to-mlp distillation framework.
\newblock In \emph{Proceedings of the AAAI Conference on Artificial Intelligence}, 9, 10351--10360.

\bibitem[{Yan et~al.(2020)Yan, Chen, Zhang, and Li}]{yan2020neural}
Yan, J.; Chen, S.; Zhang, Y.; and Li, X. 2020.
\newblock Neural architecture search for compressed sensing magnetic resonance image reconstruction.
\newblock \emph{Computerized Medical Imaging and Graphics}, 85: 101784.

\bibitem[{Yang et~al.(2022)Yang, Zhou, An, Jiang, Xu, and Zhang}]{yang2022cross}
Yang, C.; Zhou, H.; An, Z.; Jiang, X.; Xu, Y.; and Zhang, Q. 2022.
\newblock Cross-image relational knowledge distillation for semantic segmentation.
\newblock In \emph{Proceedings of the IEEE/CVF Conference on Computer Vision and Pattern Recognition}, 12319--12328.

\bibitem[{Zagoruyko and Komodakis(2017)}]{zagoruyko2016paying}
Zagoruyko, S.; and Komodakis, N. 2017.
\newblock Paying More Attention to Attention: Improving the Performance of Convolutional Neural Networks via Attention Transfer.
\newblock In \emph{The 5th International Conference on Learning Representations}.

\bibitem[{Zhang et~al.(2022)Zhang, He, Yang, Li, Wei, Huang, Zhang, He, and Zheng}]{zhang2022mmformer}
Zhang, Y.; He, N.; Yang, J.; Li, Y.; Wei, D.; Huang, Y.; Zhang, Y.; He, Z.; and Zheng, Y. 2022.
\newblock mmformer: Multimodal medical transformer for incomplete multimodal learning of brain tumor segmentation.
\newblock In \emph{International Conference on Medical Image Computing and Computer-Assisted Intervention}, 107--117.

\bibitem[{Zhao, Yang, and Sun(2022)}]{zhao2022modality}
Zhao, Z.; Yang, H.; and Sun, J. 2022.
\newblock Modality-adaptive feature interaction for brain tumor segmentation with missing modalities.
\newblock In \emph{International Conference on Medical Image Computing and Computer-Assisted Intervention}, 183--192. Springer.

\bibitem[{Zhou et~al.(2023)Zhou, Guo, Zhang, Han, Yu, Wang, and Yu}]{10183842nnformer}
Zhou, H.-Y.; Guo, J.; Zhang, Y.; Han, X.; Yu, L.; Wang, L.; and Yu, Y. 2023.
\newblock nnFormer: Volumetric Medical Image Segmentation via a 3D Transformer.
\newblock \emph{IEEE Transactions on Image Processing}, 32: 4036--4045.

\end{thebibliography}

\end{document}


\maketitle

\section{Effectiveness in Semantic Segmentation of Natural Images}
\label{natural}
To showcase the generalization capability of our method, we extend the evaluation of our proposed SMML framework to additional downstream applications. Specifically, we perform experiments on semantic segmentation of natural images using the NYUv2 dataset \cite{silberman2012indoor}.

The NYUv2 dataset contains 1,449 indoor RGB-D images, of
which 795 are used for training and 654 for testing. 
Employing the standard 40-class labeling scheme, we compare the performance of our SMML method with other approaches including the HeMIS \cite{havaei2016hemis} LCR \cite{zhou2020brain}, RFNet \cite{ding2021rfnet}, mmFormer \cite{zhang2022mmformer}, and MMANet \cite{wei2023mmanet}.
In this scenario, we adhere to the standard protocol to evaluate the semantic segmentation performance using the metric of mean Intersection over Union (mIOU), where higher mIOU values indicate better segmentation performance.

Table \ref{tab1:env} demonstrates that our method achieves the best performance for all the modality combinations.
To be specific, it outperforms the second-best method, MMANet, by 1.3\%, 0.6\%, 1.4\% in the three modality combinations, respectively.
Overall, our method enhances the MMANet's average mIOU score by 1.3\%.
This demonstrates that our SMML framework not only delivers superior performance in brain tumor segmentation but also shows promise for the incomplete multi-modal segmentation of natural images. This further validates that our SMML framework effectively handles input with arbitrary missing modalities.


\begin{table}[ht]
    \centering
    \scalebox{0.75}{
    \begin{tabular}{cc|c|c|c|c|c|c}
        \toprule
        \multicolumn{2}{c|}{Modality} & \multicolumn{6}{c}{mIOU($\uparrow$) }    \\   
        \midrule
        RGB& Depth &HeMIS &LCR &RFNet &mmFormer &MMANet&Ours   \\
        \midrule
         $\bullet$&$\circ$&33.2&41.9&42.9&43.2&44.9&\textbf{46.2} \\
         $\circ$&$\bullet$&31.2&39.9&40.8&41.1&42.8&\textbf{43.4} \\
         $\bullet$&$\bullet$&37.8&47.5&48.1&48.5&49.6&\textbf{51.0}  \\
         \midrule
        \multicolumn{2}{c|}{Average}&34.1&43.1&43.9&44.3&45.6&\textbf{46.9}  \\
        \bottomrule
    \end{tabular}}
    \caption{Comparison results on the NYUv2 dataset. }
    \label{tab1:env}
\end{table}

\bibliography{aaai25}